\documentclass[sn-mathphys-ay]{sn-jnl}% Math and Physical Sciences Numbered Reference Style 
\usepackage{graphicx}%
\usepackage{multirow}%
\usepackage{amsmath,amssymb,amsfonts}%
\usepackage{amsthm}%
\usepackage{mathrsfs}%
\usepackage[title]{appendix}%
\usepackage{xcolor}%
\usepackage{textcomp}%
\usepackage{manyfoot}%
\usepackage{booktabs}%
\usepackage{algorithm}%
\usepackage{algorithmicx}%
\usepackage{algpseudocode}%
\usepackage{listings}
\usepackage[capitalise]{cleveref}
\usepackage{hyperref}

\usepackage[T1]{fontenc}
\usepackage[thinc]{esdiff}

\DeclareMathOperator*{\argmax}{arg\,max}

\newtheorem{theorem}{Theorem}%  meant for continuous numbers
\newtheorem{proposition}[theorem]{Proposition}% 
\newtheorem{definition}{Definition}%

\begin{document}

\title[Democratic Resilience and Sociotechnical Shocks]{Democratic Resilience and Sociotechnical Shocks}


 % \author{Authors Removed for Anonymity}%\email{rahimian@pitt.edu}

 \author*[1]{\fnm{M. Amin} \sur{Rahimian}}\email{rahimian@pitt.edu}

 \author[2,3]{\fnm{Michael P.} \sur{Colaresi}}\email{mcolaresi@pitt.edu}

 \affil*[1]{\orgdiv{Department of Industrial Engineering}, \orgname{University of Pittsburgh}, \orgaddress{\street{Benedum Hall}, \city{Pittsburgh}, \postcode{15261}, \state{PA}, \country{USA}}}

 \affil[2]{\orgdiv{Department of Political Science}, \orgname{University of Pittsburgh}, \orgaddress{\street{Wesley W. Posvar Hall}, \city{Pittsburgh}, \postcode{15260}, \state{PA}, \country{USA}}}

 \affil[3]{\orgname{Peace Research Institute Oslo}, \orgaddress{Hausmanns Gate 3}, \postcode{NO-0186} \city{Oslo}, \country{Norway}}


\abstract{We focus on the potential fragility of democratic elections given modern information-communication technologies (ICT) in the Web 2.0 era. Our work provides an explanation for the cascading attrition of public officials recently in the United States and offers potential policy interventions from a dynamic system's perspective. We propose that micro-level heterogeneity across individuals within crucial institutions leads to vulnerabilities of election support systems at the macro scale. Our analysis provides comparative statistics to measure the fragility of systems against targeted harassment, disinformation campaigns, and other adversarial manipulations that are now cheaper to scale and deploy. Our analysis also informs policy interventions that seek to retain public officials and increase voter turnout. We show how limited resources (for example, salary incentives to public officials and targeted interventions to increase voter turnout) can be allocated at the population level to improve these outcomes and maximally enhance democratic resilience. On the one hand, structural and individual heterogeneity cause systemic fragility that adversarial actors can exploit, but also provide opportunities for effective interventions that offer significant global improvements from limited and localized actions.}




\maketitle

%%%%%%%%%%%%%%%%%%%%%%%%%%%%%%%%%%%%%%%%%%%%%%%%%%%%%%%%%%%%%%%%%%%%%%%%%%%%%%%%
\section{Introduction}

Social networks are critical sociotechnical infrastructures that facilitate the flow of information in liberal democracies. In-person and digitally mediated communication can spread healthy behavior and democratic engagement, but can also be used to target vulnerable communities with false and misleading content that sows discord and mistrust. Recent work has highlighted the role of social networks in spreading vaccine misinformation and violent conspiratorial content that disrupted our public health and national elections during the pandemic \citep{o2019misinformation, impact-covid19-vaccine-misinfo2022}. Such concerns are amplified when it comes to the operation of democratic societies that rely on collective decision processes to implement effective policies \citep{how-our-misunderstanding}.

Election security and public trust in voting results have received unprecedented attention over the last several years in the US and abroad \citep{protecting-elections-from-social-media-manipulation,securing-US-election-systems,election-security}. At the domestic level, the US Department of Homeland Security identified election systems as a critical domestic security infrastructure in 2017. This prioritization highlighted the need to protect the institutions, personnel, hardware and software used to tabulate, audit, and communicate votes, from storage facilities and polling places to voter registration databases and voting IT infrastructure \citep{us1starting,national2018securing}. At the macro-level, vigorous challenges from autocratic regimes and non-liberal ideologies have re-emerged and triggered public concern about the future strength and viability of democratic institutions and liberal values moving forward \citep{the-return-of-the-presidential-putsch}. 
Furthermore, due to the centrality of the United States in international affairs and in particular NATO, the resilience and fragility of specific US democratic institutions have global implications \citep{remarks-by-president-biden}. Thus, how to identify and fix vulnerabilities in the human and technological infrastructure that US elections depend on, in the face of domestic and international pressure, is of preeminent importance. 

% Elections are highly decentralized in the U.S., with over $80,000$ election officials in $10,000$ election offices facilitating voting nationwide\cite{election-deniers-already-are-disrupting-the-midterm-election,election-officials-testify-about-threats-and-election-security,poll-of-local-election-officials,election-administration-at-state-and-local-levels}.

Elections in the United States also have several unique facets, most notably their decentralized character, that shape potential sources of vulnerability. There are over $80,000$ election officials across $10,000$ offices working on various aspects of national election voting, interpreting, and acting on different statutes at the local and state levels \citep{election-deniers-already-are-disrupting-the-midterm-election,election-officials-testify-about-threats-and-election-security,poll-of-local-election-officials,election-administration-at-state-and-local-levels}. Despite their integral role in election security, officials are leaving their election roles, with some reports that up to a third of election officials do not return in some states \citep{election-officials-are-burnt-out-and-leaving,BipartisanCenterReportData2024}. Although there are a variety of potential explanations for this trend, including tension over pandemic voting procedures and an aging workforce, there are numerous reports of increased and acute harassment of election officials from other citizens across the country \citep{BipartisanCenterReportData2024}. Based at least in part on unsubstantiated claims of election tampering from President Donald Trump, some citizens have aggressively accused election officials of improperly administering elections, including the 2020 presidential election \citep{facing-harassment-and-death-threats}. This harassment has often taken the form of doxxing, as well as online and offline threats and intimidation of local election officials \citep{Reuters}. However, scholars have yet to systematically model the effects of this harassment of election officials. 

Research has underlined the role that social media platforms and their sociotechnical affordances play in online harassment.\footnote{While we focus on social media here, online harassment is enabled by a host of current information communication technologies, including the mass adoption of smartphones that connects people to their avatars regardless of location, search engines that allow election officials to be identified, and provides many of the documents that are used in harassment, and browsers and ad technology that allow tracking and the targeting of individuals based on their past behavior and imputed values and characteristics.} Online disinformation is influencing citizens' judgments about fairness of elections, proliferating conspiracy theories, and increasing the anxiety of internet users about whether the democratic nature of the country can be guaranteed \citep{youtube-the-great-radicalizer,qAnon-conspiracy-theories-and-social-media-warfare,internet-trolls-against-russian-opposition}. In addition, the proliferation of bots on social networks can drive the spread of disinformation, which affects people's perception of the integrity of elections \citep{presidential-elections-in-ecuador,donovan2022meme}. 

This research has led to suggested steps to mitigate the risks of threats to democracy from social media, including flagging or limiting the dissemination of suspect information \citep{social-media-and-misleading-information-in-a-democracy,multi-Platform-analysis,shifting-attention-to-accuracy-can-reduce-misinformation-online}, identifying inauthentic accounts \citep{towards-proactive-moderation-of-malicious-content-via-bot-detection-in-fringe-social-networks}, and building computational models of adversarial exploits \citep{bipartite-consensus-of-opinion-dynamics-through-delivering-credible-information,an-adversarial-model-of-network-disruption}. Although these methods do alleviate some of the online harassment, harassment is not only online, election officials may still be subject to the offline harassment mentioned above. In this paper, our aim is to model harassment from a systemic perspective with a more inclusive conception of harassment that includes harassment that occurs offline in board rooms and on street corners.  

%Each official is endowed with a threshold that is a random draw from a distribution of thresholds. The variability in the distribution of thresholds represents 

% In the following sections, we address this issue by developing a model for attrition of election officials facing harassment. We find that there exists a maximum harassment that the system can tolerate that depends on the distribution  and heterogeneity of individual tolerances. Following the basic model, we alter some of our assumptions to investigate how the addition of' partisan election officials, who do not value or contribute to the running of high-quality elections, affects the model's findings. We find that partisan election officials replacing non-partisan ones can result in worse conditions for the election system. We then explore the effects of potential policy interventions in our model, specifically increasing election official pay. Inspired by ideas from the Optimal Transport Problem (OTP), we investigate the consequences of varying the number of payments to officials by computing the distances between vertices and finding the optimum intervention to maximize resilience to harassment subject to the costs of the payments.

\subsection*{Our Contribution}
 % We model the dynamics of attrition of election officials who leave their posts in the face of harassment and adverse work conditions. Each official is endowed with a threshold that is a random draw from a distribution of thresholds. The variability in the distribution of thresholds represents the heterogeneity in the officials' tolerance towards harassment, their dedication to the election system, and their varying work conditions, e.g., salary and benefits. As lower-endowed officials leave the system, the remaining officials face even higher pressure from the increased workload and more targeted harassment, and more attribution ensues. We propose to characterize the resiliency of the election system as the proportion of officials that remain with the systems in the face of a fixed or time-varying harassment schedule and identify the factors that influence this resiliency, including threshold heterogeneity and distributional features, the structure of support networks among the officials, and the state's policy in replacing officials who leave. Our results with continuous distributions reveal the existence of a critical harassment level, $H^*$, beyond which the election systems cannot survive and this critical value is directly related to the dispersion of the distribution of thresholds. The latter can be modified by offering incentives (e.g., increased salaries) to officials, which we propose to model as an optimal (least cost) transport problem from the existing distribution of threshold to a more resilient one in Section \ref{sec:OTP}. 

We develop a dynamic model that allows for the attrition of election officials who leave their posts in the face of harassment. The heterogeneity in the officials' tolerance towards harassment represents the variability in their work conditions, dedication to the elections system, as well as their connections to the community and civic networks. As lower-endowed officials leave the system, the remaining officials face higher pressure from the increased workload and more targeted harassment. We propose characterizing the resiliency of the election system as the proportion of officials that remain with the systems in the face of a fixed or time-varying harassment schedule and identify the factors that influence this resilience, including the structure of support networks among the officials and the system's ability in replacing officials who leave. We also explore how plausible sociotechnical shocks that lower the costs for harassers and/or allow for targeting of specific sets of officials for maximum disruptive effect. Our results reveal the existence of a critical harassment level beyond which an elections system cannot survive, and this critical value is directly related to the dispersion of the distribution of thresholds. The latter can be modified by offering incentives (e.g., increased salaries) to officials, which we propose to model as an optimal (least cost) transport problem from the existing distribution of thresholds to a more resilient one. The factors and metrics that we identify with this analysis can also be quantified with data collected from administrative records, social media, or survey instruments.

 In the simplest model, we try to isolate the effect of distributional features as a factor. Each official can tolerate a specific threshold of harassment, modeled as an independent draw from a cumulative distribution function $F$, and once their harassment exceeds that threshold, they leave the community. In this model, we can measure the asymptotic proportion of election officials who remain in the community as a function of the distributional parameters of the thresholds. To simplify the model, we consider a continuum of officials that spans the unit interval $[0,1]$. At $t=0$, a fraction $F(H)$ whose thresholds are less than $H$ leave, and the remaining fraction $1-F(H)$ will face amplified harassment $H/(1-F(H))$, hence, at $t=1$ more officials leave and the fraction remaining at $t = 1$ is given by $1 - F(H/(1-F(H)))$, so on and so forth. Let $p_t$ denote the fraction that is missing from the unit interval at time $t$, then the dynamics of attrition on the unit interval can be expressed as $p_{t+1} = F({H}/(1 - p_t)), t = 0,1,2\ldots$. Let $p_{\infty}(H) = \lim_{t\to\infty} p_t$ be the asymptotic proportion that eventually leaves; the process continues until there are no officials left: $p_{\infty}(H)=1$, or all remaining officials have high enough thresholds to tolerate net harassment $H$: $p_{\infty}(H) = F({H}/{(1 - p_{\infty}(H))})$. 

\subsubsection*{A Systemic Resilience Metric} 
\begin{figure}
\centering   
\includegraphics[width=0.5\textwidth]{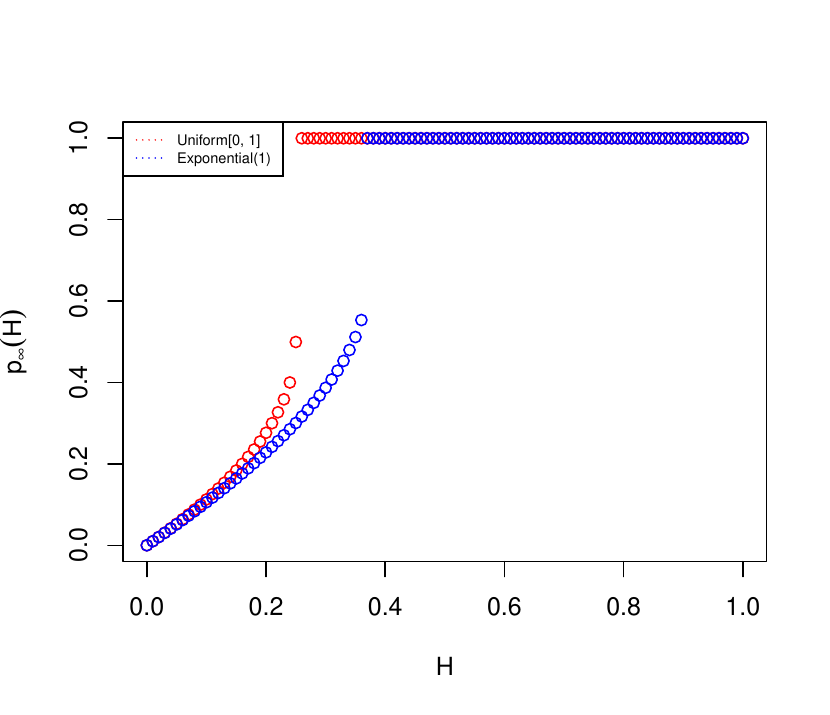}
\caption{{\scriptsize We plot $p_{\infty}(H)$ as a function of $H$ for $F_1 \sim \mbox{Uniform}[0, 1]$ in red, $F_2 \sim \mbox{Exponential}(1)$ in blue, whose corresponding resilience are $1/4$ and $1/e$, respectively. Note the stochastic dominance relationship between these distributions: $F_2 \succ F_1$ which is reflected in their resilience metrics:   $\mathcal{R}(F_2) =  {1}/{e} > \mathcal{R}(F_1) = {1}/{4}$. We can compute $\mathcal{R}(F)$ explicitly as a function of distribution parameters, for example, $\mathcal{R}(F) = b^2/(4(b-a))$ for $\theta \sim \mbox{Uniform}[a, b]$, whereas for exponential distribution $\mathcal{R}(F) = 1/(e \lambda)$ for $\theta \sim \mbox{Exp}(\lambda)$. One may look into the entirety of the $p_\infty(H)$ plot and use more nuanced statistics to better distinguish different distributions. For example, the area under the curve (AUC) of $p_\infty(H)$ over a fixed range, $[H_{\min},H_{\max}]$, can give a more refined characterization of resilience and be used as an alternative metric: $\int_{H_{\min}}^{H_{\max}}p_\infty(H)dH$. One particular value of interest would be $p_{\infty}(\mathcal{R}(F))$: the fraction remaining in the system at the critical harassment value $\mathcal{R}(F)$.} \vspace{-5pt}}
\label{fig:distributions}
\end{figure}

Note that $p_t =  1$ is always a fixed point of the $p_{t+1} = F({H}/(1 - p_t))$ dynamics, since $\lim_{x\to\infty}F(x) = 1$. In fact, we can show that the sequence $\{p_t, t = 0,1,2, \ldots\}$ is bounded and increases monotonically and has a unique sequential limit $p_{\infty}$. When $p_{\infty} = 1$, the election support system suffers a catastrophic failure: \textit{a total unraveling of the election support system}. Therefore, a key metric for systemic resilience is the smallest amount of input harassment that unravels the election support system: $\mathcal{R}(F) = \inf \{H>0 : p_{\infty}(H) = 1\}$, and it can be characterized as a function of distributional parameters. We can show that when the cumulative distribution function, $F$, has an inverse, i.e., its quantile function $Q = F^{-1}$ exists and is unique, then $\mathcal{R}(F) = \sup_{p \in (0,1)} (1-p)Q(p)$. This metric satisfies natural properties, for example, when two distributions are stochastically ordered such that $F_2$ stochastically dominates $F_1$, $F_2 \succ F_1$ --- i.e., $F_2(\theta) \leq F_1(\theta)$ for all $\theta$ with strict inequality for at least one value of $\theta$, then $\mathcal{R}(F_2) \geq \mathcal{R}(F_1)$. Figure \ref{fig:distributions} shows the plots of $p_{\infty}(H)$ versus $H$ for different distributions with discontinuities occurring at $H = \mathcal{R}(F)$. 
% The continuum analysis has the benefit of analytical tractability and being amenable to closed-form optimal transport solutions when considering optimal intervention designs in \textcolor{red}{Section} \ref{sec:OTP}. %However, our analysis of sampling distributions indicate important differences between the continuum asymptotic behavior and the large sample limits of the dynamics on finite populations. Our results in this task should address this gap to derive precise, interpretable and practically relevant notions of resilience for different threshold distributions.

\subsubsection*{Risks and Opportunities with Sociotechnical Shocks} 
In the baseline model presented above, the amount of total harassment ($H$) targeted at the election administrators is fixed and equally distributed between the election officials who remain with the system. This situation is analogous to pre-social-media harassment occurring at board meetings attended by all officials. Where in the past harassment faced by election officials during a board meeting would have been tolerable, new information-communication technologies, such as social media, make it easy for an adversary to produce harassment cheaply, at a scale that approaches or exceeds our characterization of resilience metric $\mathcal{R}(F)$, putting the election support system at risk of catastrophic unraveling. Social media surveillance and targeting technologies further enable adversaries to focus their harassment on the most at-risk individuals. For example, when thresholds are perfectly observable, an adversary's optimum strategy at each time step is to target the maximum threshold of officials such that after distributing the net harassment $H$ among the targeted group, they all leave.

In \Cref{sec:related-work}, we provide detailed related works to contextualize our results. We describe the attrition model and our proposed resilience metric in \cref{sec:attrition}. Next in \cref{sec:inflitration} we consider the effect of new recruit measures and the increased fragility of replacing officials with rogue agents. In \cref{sec:surviellence,sec:plan}, we consider the cases where harassment can be targeted in the population or planned over time. In \cref{sec:intervention}, we consider interventions to improve resilience at the lowest cost. We provide discussion and conclusion remarks in \cref{sec:conc}. Proofs and mathematical details are in the appendix.  

\section{Related Work}\label{sec:related-work}

Research on election security has focused mainly on infrastructure and not people, with a focus on voting machines \citep{securing-US-election-systems, election-security,discussion-of-election-security,experts-warn-of-dangers-from-breach-of-voter-system-software,specter2021security,barretto2021improving,wolchok2010security,appel2020ballot, halderman2022antrim}, ballot drop box locations \citep{schmidt2024drop}, and polling locations \cite{schmidt2022designing}. An exception has been the study of waiting times in polls \cite{schmidt2022designing} and the ease of use of mail-in ballots \cite{schmidt2024drop}.  However, in addition to voters and infrastructure, local election officials are critical to the operation of trusted elections \citep{election-officials-testify-about-threats-and-election-security,poll-of-local-election-officials,election-administration-at-state-and-local-levels}. 

Public administration researchers and those focused on sustainable infrastructure have highlighted the key operational role local election officials play in elections. In addition, this work focuses on internal motivations of public sector employees that we model by individual thresholds. Election officials are presumed to share a largely intrinsic and prosocial public service motivation (PSM) \citep{perry1990,haider2019}. PSM has been theorized as a relatively stable characteristic in an individual \citep{wright2010}; however, PSM has also been shown to decline under conditions such as perceived bureaucracy and red tape \citep{moynihan2007}, ``reality shock'' \citep{schott2019}, and particularly in extreme conditions such as warfare \citep{braender2013}. Our work therefore can be seen as examining the PSM of local election officials (LEOs) in the context of currently high attrition---up to a third in some states \citep{election-officials-are-burnt-out-and-leaving}. In addition, our model places this crucial election infrastructure into the current atmosphere of extreme partisanship, affective polarization, and harassment from fellow citizens \citep{facing-harassment-and-death-threats,Reuters,brennancenter2023,so2021,watters2022}. Although loss of experience and knowledge of one individual may have little effect on a given election, especially in settings such as Pennsylvania, where election governance is bipartisan, more knowledgeable and better motivated electoral workers have been shown to exert better technical performance as an organization \citep{james2019}, i.e., to perform better elections. Patterns of harassment against officials by members of the public may have deleterious effects not only on the psychology of individual officials \citep{so2021,brennancenter2023}, but may also erode organizational, technical, and administrative performance and eventually public trust in elections and voter turnout \citep{alvarez2008}. In their analysis of systemic risk in the voting supply chain, from the point where the ballots are cast to the counting and reporting of election results, \citet{scala20,scala2021proactive,scala2022evaluating,scala22,dehlinger2021poll} find insider threats -- the risk of rogue agents -- to be the most significant factor and offer ways to mitigate and improve the resilience of the election infrastructure accordingly.  

Despite the potentially serious consequences of election harassment and the interdependent dynamics that can lead to the unraveling of fundamental democratic institutions; to the best of our knowledge, most existing literature only tangentially engages with these potential consequences, the specific role of technological shocks, and frameworks to rebuild resilience.  One of the closest papers, by \citet{dbt2006}, discusses the bribing and harassment of public officials as a two-stage game in which a population of prospective officials with heterogeneous qualities decides to apply for public office or enter the private sector. In the second stage, the interaction of public officials with pressure groups determines the amount of public good provided. The authors investigate the equilibrium effects of bribes and threats and the consequences of policies such as official immunity. Other research focuses on strategic behaviors and incentives within election systems: \citet{LEONTIOU2023471} illustrate bandwagons in large and especially close elections, \citet{ANDONIE2012681} propose a two-stage game for opinion polls followed by elections, and find that the unique equilibrium for competing parties relates to their differing sample sizes. In empirical work, \citet{BERMAN2019292} discuss the fairness of elections in Afghanistan and \cite{WOON2019735} investigate a model of elections with candidate honesty and present experimental results using task performance as a laboratory model of policy making. In other works, \citet{APFFELSTAEDT2022148} and \citet{LEMOGLIE2019140} look at the effects of elections on changing social norms and the news coverage of negative behaviors such as corruption, which is related to, but distinct from, targeted harassment that straddles the online and offline worlds.

Our work also relates to studies of attrition and retention in the literature on labor economics, human resources, and organization science. The microeconomic models of attrition and job search focus on the match between the firm and workers given the surplus generated from employment and their outside options \citep{laing2011labor}. Attrition can occur due to incomplete information about the surplus or the quality of the match or by search when the expected utility of outside options exceeds the surplus of remaining in employment \citep{mortensen1999new}. Market outcomes vary according to job characteristics and population characteristics. Attrition is also of interest to management and organization sciences where significant focus has been devoted to predictors of attrition such as overwork \citep{holtom20085}, and the negative impact of attrition on organizational performance, e.g., lowering profits and increasing delays \citep{park2013turnover,narayanan2009matter}. The public service of election institutions and the potential role of external antidemocratic adversaries provide unique contours for our current contribution. 

\section{Attrition and Civic Unraveling}\label{sec:attrition}

We will motivate our model in the context of election systems, but the main conclusions apply broadly to civic networks when faced with harassment pressure. In our harassment model, election officials are faced with total harassment of $H$ in their community, the local civic network with election officials and voters. Political participation is costly \citep{an-economic-theory-of-political-action-in-a-democracy,the-pure-theory-of-large-two-candidate-elections}, and voters prefer to participate in high-quality elections \citep{perceptions-of-electoral-integrity-in-consolidating-democracies}. They rely on election officials for the administration of their elections, for information dissemination related to the election, and broadly to ensure the integrity and security of their votes. Each official has a tolerance for a specific threshold of harassment that also depends on the support that they receive from other officials in the system. Once their tolerance for harassment is exceeded, the officials will leave--they will permanently vacate their civic post. We consider a continuum of agents (civic network participants) indexed by the unit interval $\mathcal{I} = (0,1)$. We microfound the agents' decisions to stay within the support network as follows. Each agent $i \in \mathcal{I}$ is endowed with a payoff-relevant type $\theta_i$ that is an independent draw from a common distribution $F$. A high type represents a higher public service motivation, and the heterogeneity in $\theta_i$ values translates into the agents' differing tolerances for harassment. In particular, we assume that each agent $i$ at time $t$ has the option to stay in the network $a_{i,t} = 1$ or leave $a_{i,t} = 0$. Given a harassment level $H_t$ and assuming zero utility for the outside option $a= 0$, we can formulate the utility of agent $i$ at time $t$ as follows: 
\begin{align}
U_{i,t}(a) = a(\theta_i -H_t/(1-p_t)), a \in \{0,1\}, \label{eq:utility}    
\end{align} where $0\leq  p_t \leq 1$ is the fraction (Lebesgue measure on the unit interval) of agents who have left the system at time $t$, leaving the remain $(1-p_t)$ fraction to experience $H_t/(1-p_t)$ harassment equally spread. The utility maximizing action in \cref{eq:utility} implies a threshold on the ``experienced" harassment, $H_t/(1-p_t)$, such that:
\begin{align}
a_{i,t} = \argmax_{a\in\{0,1\}} U_{i,t}(a) = \left\{
    \begin{array}{lr}
        0, & \text{if } \theta_i \leq H_t/(1-p_t),\\
        1, & \text{otherwise.}
    \end{array}
\right. \label{eq:threshold-behavior}
\end{align} 
To interpret the utility function in \cref{eq:utility} and the resultant threshold behavior in \cref{eq:threshold-behavior}, we assume that the total amount of harassment input to the system is equally distributed so that as agents leave the civic network, the harassment experienced by the remaining agents increases proportionally as $H_t/(1-p_t)$. The remaining agents tolerate their increasingly experienced harassment until their threshold is exceeded, $\theta_i \leq H_t/(1-p_t)$, at which point they leave the network $a_{i,t} = 0$. This situation is analogous to pre-social-media harassment occurring at board or other public meetings attended by all officials, with the targeting being general and diffuse. Similarly, when either general or specific allegations appear online the board confers together, paying a cost in time and effort. Finally, there are many cases where the board is jointly named as plaintiffs in complaints \citep{BipartisanCenterReportData2024}.\footnote{See specific examples at \url{https://elections.wi.gov/resources/complaints/el-24-87-harbridge-v-lundgren} and \url{https://www.pa.gov/en/services/vote/file-an-election-complaint.html}.} The assumption that the total harassment $H_t$ is equally spread among the remaining $(1-p_t)$ officials, and therefore experienced as $H_t/(1-p_t)$, provides a reasonable baseline for our characterization of resilience in \Cref{sec:systemicresiliencemetric}. In \Cref{sec:surviellence}, we investigate the implications of targeting harassment and the reduced resilience that results from having a subset of officials negatively affected. 

An alternative formulation of the utility that leads to the same threshold behavior as \Cref{eq:threshold-behavior} is $U_{i,t}(a) = a(\theta_i(1-p_t) -H_t)$. In this formulation, the payoff-relevant type $\theta_i$ measures the positive externality of remaining within the civic network, given that there are $(1-p_t)$ remaining officials. Agents receive more externality from remaining in larger networks and are therefore more likely to tolerate experienced harassment $H_t$. As more agents leave the civic network, their positive externality is lost on the remaining agents, which in turn causes more departures and threatens a civic unraveling of the support network. 

The dynamics of subsequent unraveling is determined by the distribution $F$, and we can use the distribution parameters to investigate the asymptotic fraction of agents that remain with the system. 

Initially ($t=0$), the election official will be removed from the community if harassment $h_0 = H$ exceeds their threshold. Due to the heterogeneity of the threshold, some proportion of officials $p_0$, for whom $h_0$ is the quantile, will leave the community. Specifically, $p_0 = F(h_0)$, where $F$ is the cumulative distribution function (CDF) of the thresholds. 
Once $p_0$ of the officials leave, the harassment experienced by each of the remaining officials increases. To be specific, at time $t=1$, the harassment experienced by the officials is $h_1=\frac{H_0}{1-p_0}$. Because the experience of harassment is greater, a higher proportion of officials will leave ($p_1$), which can also be expressed in terms of the CDF as follows: $p_1 = F(H/(1-p_0) = F(H/(1-F(H)))$. Let $p_{\infty}(H) = \lim_{t\to\infty} p_t$ be the proportion when the process converges under $H$: the process will continue until no officials leave ($p_{\infty}(H) < 1$), or until all officials have left ($p_{\infty}(H) = 1$). The dynamic and the equilibrium can be expressed as follows:
\begin{align}
    p_{t+1} &= F(\frac{H}{1 - p_t}), \label{eq:model-1-sys-dynamics} \\
     p_{\infty}(H) &= F(\frac{H}{1 - p_{\infty}(H)}). \label{eq:model-1-sys-equilibrium}
\end{align}

Given $H> 0$, the process starts with $p_0 =F(H)$. The support of $\theta$ is any positive number $(0, \infty)$ so $F(x) = 0, \forall x \leq 0$. Note that $p_t \in [0,1]$ and $p_{\infty}(H) = 1$ is always a fixed point of \eqref{eq:model-1-sys-dynamics}. The following observation (for completeness, a proof is provided in \cref{app:monotonicity-convergence}) justifies the asymptotic analysis of \cref{eq:model-1-sys-dynamics} dynamics to relate $p_{\infty}(H)$ to properties of the distribution $F$:
\newline

% Based on the previous models, we attempt to find the relationship between the distribution of election officials' threshold and $p_{\infty}(H) = \lim_{t\to\infty} p_t$ in the model through asymptotic analysis, in order to analyze the robustness of the system of different types of distributions.

% Before analyzing the limiting values of $P_t$, we perform certain transformations of the variables to simplify the symbols' expressions and define the domain of variables to ensure their meaningfulness. Notice that:
% \begin{align}
% P_{t+1} & =F\left(\frac{H}{1-P_t}\right) \\
% & =\mathbb{P}\left(\theta \leqslant \frac{H}{1-P_t}\right) \\
% & =\mathbb{P}\left(1-P_t \leqslant \frac{H}{\theta}\right) \\
% & =\mathbb{P}\left(1-\frac{H}{\theta} \leqslant P_t\right)
% \end{align}
% Here, $\theta$ is the election officials' threshold to withstand harassment. For a given election official in the system, the threshold must be greater than zero, otherwise, the election official will leave the election system regardless of the presence of harassment in the system. Therefore, $\theta \in(0, \infty)$. Denote that $\varphi=1-\frac{H}{\theta}$, then we have:
% \begin{align}
% P_{t+1} & =F\left(\frac{H}{1-P_t}\right) \\
% & =\mathbb{P}\left(\varphi \leqslant P_t\right) \\
% & = G(P_t)
% \end{align}
% Here, $G$ is the distribution of the transformed variable $\varphi$. In the following subsections, we prove the existence of the limit of $P_t$, which is the maximum proportion of election officials that leave the system.

\begin{proposition}[Monotonicity and Convergence]
The sequence $\{p_t\}$ is bounded and increases monotonically and therefore converges to a limit. 
\end{proposition}

\subsection{A Systemic Resilience Metric}\label{sec:systemicresiliencemetric}

Based on \cref{eq:model-1-sys-equilibrium}, we can explore two questions: (1) What is the maximum total amount of harassment $H^*$ that the community can tolerate without all election officials leaving? (2) What is the fraction of election officials who leave, $p(H^*)$, if the total amount of harassment is $H^*$? This gives rise to the following definition of resilience:

\begin{definition}[Resilience]\label{def:res}
The resilience of a community with a threshold distribution, $F$, is defined as $\mathcal{R}(F) = \inf \{H>0 \mid p_{\infty}(H) = 1\}$.
\end{definition}

\vspace{5pt}

It is clear that resilience is determined by the distribution function $F$. Note that for every $H > 0$, $p_{\infty}(H) = 1$ is always a fixed point of \eqref{eq:model-1-sys-dynamics}. Hence, if there are no roots in the interval $[0, 1)$, then $p_{\infty}(H) = 1$. On the other hand, if there exist roots that are strictly less than $1$, the sequence $\{p_t\}$ under $H$ must converge to the smallest root.  Therefore, the goal is to find the smallest $H$ such that \eqref{eq:model-1-sys-dynamics} has no root in the interval $[0, 1)$. The following theorem (proved in \cref{app:quantile}) gives us an expression of resilience $R(F)$ in terms of the quantile function $Q$ of $F$:\newline 

\begin{theorem}[Resilience and Quantile Function]\label{thm:expression} If there exists a quantile function $Q = F^{-1}$, then the resilience of $F$ can be expressed in terms of $Q$ as follows: $\mathcal{R}(F) = \sup_{p \in (0,1)} \{(1-p)Q(p)\}$.
\end{theorem}

\vspace{5pt}

\cref{fig:distributions} shows examples of $\mathcal{R}(F)$ for two common distributions where we have used the above theorem to calculate $\mathcal{R}(F)$ in closed-form in terms of the distribution parameters.

% \textcolor{red}{Fogure howed example}
% \begin{figure}[h]
%     \centering
%     \includegraphics[width=0.5\textwidth]{figs/022123.png}
%     \caption{Illustration of three different distributions with the same Resilience. The distributions  are $ Unif[0, 4/e], Exp(1), Pareto(\theta_m = 1/e, \, \alpha = 1) $ respectively. As a result, the Resilience $\mathcal{R} (F)$ may not be the only metric for the robustness of a network.}
%     \label{fig:same-Hmax}
% \end{figure}

% \subsection{\color{red}{Unravelling of election system with induction of partisan officials (Model 2)} TBD}

\vspace{5pt}

% \subsubsection*{A Stochastic Ordering}%\label{sec:stochstic-ordering}
Recall that given two distributions of thresholds whose cumulative distribution functions are $F_1$ and $F_2$, we say that $F_2$ stochastically dominates $F_1$, if for all threshold values $\theta$, $F_2(\theta) \leq F_1(\theta)$ with inequality strict for at least one value of $\theta$. Second-order stochastic dominance is a relaxation of this definition that only
requires $\int_{-\infty}^\theta\left[F_1(t)-F_2(t)\right] d t \geq 0$ for all $\theta$ with strict inequality in some $\theta$. First-order stochastic dominance implies second-order dominance. Our following result shows the implication of stochastic ordering of the threshold distribution on the maximum tolerable harassment (proved in \cref{app:stochastic-ordering}). \newline

\begin{proposition}[A Stochastic Ordering]\label{prop:stochstic-ordering} Given two distributions of thresholds $F_1$ and $F_2$, if $F_1$ stochastically dominates $F_2$ (first-order, stochastic dominance), then distribution $F_{1}$ can tolerate more harassment than distribution $F_{2}$ meaning that $\mathcal{R}(F_1) \geq \mathcal{R}(F_2)$.
\end{proposition}

\section{Recruitment and Infiltration} \label{sec:inflitration}

In this section, we consider the possibility of recruiting new officials into the system. The local administration may only partially succeed in replacing the officials who have left, and new hires may lack organizational experience and knowledge to fully replace the population that has left. We model the combined effect of these factors by a coefficient $0 \leq \alpha \leq 1$, such that if the local election administration is completely effective in replacing the leaving official, then $\alpha = 0$; and if they fail completely, then $\alpha =1$. In the latter case, we recover the previous dynamics in \Cref{eq:model-1-sys-dynamics}. The attrition dynamics with recruitment factor $\alpha$ is given by: 
\begin{align}
    p_{t+1} = F(\frac{H}{1 - \alpha p_t}).  \label{eq:model-recruitment-sys-dynamics}
\end{align} If $\alpha = 0$, then $p_t = F(H) = p_{\infty}(H)$ for all $t$ and $\mathcal{R}(F) = \theta_{\max} = \inf\{\theta>0|F(\theta) = 1\}$ if support of $F$ is bounded, and $\mathcal{R}(F) = +\infty$ otherwise. For general $\theta$ we can repeat the steps in \cref{app:quantile} to obtain: \newline 

\begin{proposition} \label{prop:model-recruitment-resilience}
Assuming there exists a quantile function $Q = F^{-1}$, the resilience metric for the recruitment-augmented model of attrition dynamics is given by $\mathcal{R}_{\alpha}(F) = \sup_{p \in (0,1)} \{(1-\alpha p)Q(p)\}$.  
\end{proposition} 

\vspace{5pt}

Although new recruitment can limit the effect of harassment and increase resilience, we can also imagine a nefarious angle to recruitment: Consider ``rogue'' agents infiltrating the election system to replace leaving officials at each time step $t$, motivated, for example, by anecdotes of election deniers occupying such positions of influence over local elections \citep{when-an-election-denier-becomes-an-election-chief}. We model the effect of rogue agents that replace the departing proportion $p_t$ as an increase in total harassment $H$ proportional to vacancies $p_t$ that can be abused by rogue agents who are not themselves targets of harassment, for example, because they are seen as sympathetic to the harassment goals \citep{the-origins-and-consequences-of-affective-polarization-in-the-united-states}. In this case, let us denote the total harassment in the community at time $t$ by $H_{t} = H_0+\lambda p_t$, where $\lambda \geq 0$ measures the increasing rate of harassment from introduction of rouge agents proportionally to $p_t$, leading to the following attrition dynamics:
\begin{align}
    p_{t+1} = F(\frac{H_{t}}{1 - p_t}) = F(\frac{H_0+\lambda p_t}{1 - p_t}).  \label{eq:model-inflitration-sys-dynamics}
\end{align} If $\lambda = 0$, then we recover the original attrition dynamics in \cref{eq:model-1-sys-dynamics}. Larger $\lambda$ indicates the growing adversarial capacity to abuse the vacancies left by the departing officials. Following the same steps as in \cref{app:quantile}, we can show the following characterization of resilience for \cref{eq:model-inflitration-sys-dynamics}: $\mathcal{R}_{\lambda}(F) := \sup_{p \in (0,1)} \{(1-p)Q(p) - \lambda p\} < \sup_{p \in (0,1)} \{(1-p)Q(p)\} = \mathcal{R}(F)$. \newline 

\begin{proposition} \label{prop:model-2-resilience}
Assuming there exists a quantile function $Q = F^{-1}$, then the resilience metric for the infiltration model of attrition dynamics is given by $\mathcal{R}_{\lambda}(F) = \sup_{p \in (0,1)} \{(1-p)Q(p) - \lambda p\}$.
\end{proposition}

\vspace{5pt}

To the extent that the local election administration is successful in replacing their departing workforce in \cref{eq:model-recruitment-sys-dynamics}, this also reduces the opportunities for adversaries to abuse vacancies from $\lambda p_t$ in \cref{eq:model-inflitration-sys-dynamics} to $\lambda \alpha p_t$; hence, combining \cref{eq:model-recruitment-sys-dynamics} and \cref{eq:model-inflitration-sys-dynamics} we obtain a joint attrition dynamic that accounts for recruitment and infiltration as follows:  
\begin{align}
    p_{t+1} = F(\frac{H_0+\lambda \alpha p_t}{1 - \alpha  p_t}),  \label{eq:model-recruitmnet-inflitration-sys-dynamics}
\end{align} \cref{eq:model-recruitmnet-inflitration-sys-dynamics} generalizes \cref{eq:model-recruitment-sys-dynamics} and \cref{eq:model-inflitration-sys-dynamics}, and we recover the prior models by setting $ \lambda = 0$ and $\alpha = 1$, respectively. The associated resilience metric is given by: \newline

\begin{proposition} \label{prop:model-inflitration-recruitment-resilience}
Assuming there exists a quantile function $Q = F^{-1}$, the resilience metric for the dynamics of attrition with recruitment and infiltration is given by $\mathcal{R}_{\alpha,\lambda}(F) = \sup_{p \in (0,1)} \{(1-\alpha p)Q(p) - \lambda\alpha p\}$.
\end{proposition} 

\vspace{5pt}

\begin{figure}
\centering   
\includegraphics[width=0.7\textwidth]{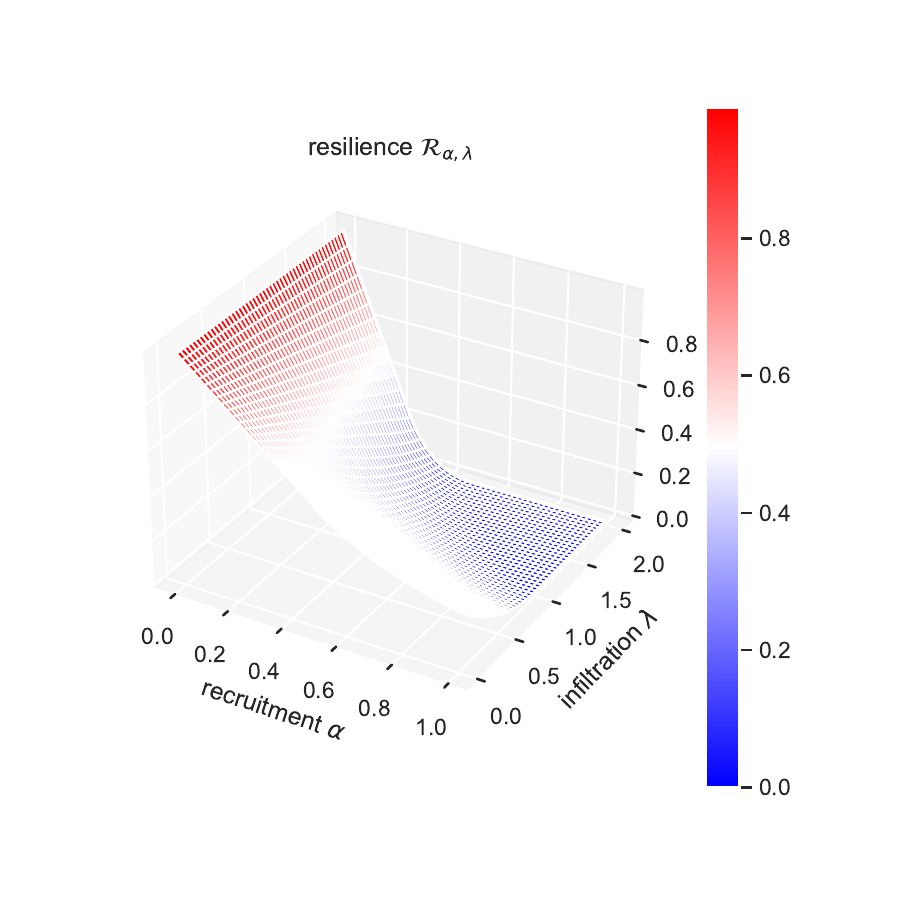}
\vspace{-20pt}
\caption{{\scriptsize We plot $\mathcal{R}_{\alpha,\lambda}(\mbox{Uniform}[0, 1])$ as a function of recruitment parameter $\alpha$ and infiltration $\lambda$. As $\alpha \to 0$, the local administration is perfectly able to replace departing officials and resilience converges to its maximum value of one corresponding to the upper support of the $\mbox{Uniform}[0, 1]$ threshold distributions. As $\alpha \to 1$ and for larger values of $\lambda$ the resilience is zero; the system becomes completely fragile and faces complete unraveling with any input harassment $H>0$.} \vspace{-5pt}}
\label{fig:resilience}
\end{figure}

We can calculate the resilience metric $\mathcal{R}_{\alpha,\lambda}(F)$ for the examples in \cref{fig:distributions} to see the effect of the recruitment parameter $\alpha$ and infiltration $\lambda$. For $F_1 \sim \mbox{Uniform}[0, 1]$, we have (\Cref{fig:resilience}):
\begin{align}
\mathcal{R}_{\alpha,\lambda}(F_1) =
\left\{
	\begin{array}{ll}
		1 -\alpha(1+\lambda) & \mbox{if }  \alpha \leq 1/(2+\lambda)\\
		(1-\lambda \alpha)^2/(4\alpha) & \mbox{if } 1/(2+\lambda) \leq \alpha \leq 1/\lambda\\
        0 & \mbox{if } 1/\lambda \leq \alpha 
	\end{array}
\right.
\end{align}

The situation for $F_2 \sim \mbox{Exponential}(1)$ is very different. If the local administration can recruit new officials to any degree of effectiveness, that is, $\alpha<1$, then $\mathcal{R}_{\alpha,\lambda}(F_2) = +\infty$ for all $\lambda>0$ and $\alpha<1$. If $\alpha = 1$, that is, the local administration is completely ineffective in replacing departing officials, then $\mathcal{R}_{1,\lambda}(F_2) = e^{\lambda-1} -\lambda$ for $0\leq\lambda\leq1$ and $\mathcal{R}_{1,\lambda}(F_2) = 0$ for $\lambda >1$. Having officials who can tolerate arbitrarily large harassment (even if they constitute an exponentially small tail of the population, as is the case of $F_2 \sim \mbox{Exponential}(1)$) makes the ability of the local administration to replace departing officials ultimately important to achieve resilient election infrastructure.  %any degree of adversarial infiltration is enough to render the system completely fragile

\section{Surveillance and Targeting}\label{sec:surviellence}

Consider a sociotechnical shock that enables adversaries to target at-risk officials at every time step, that is, the adversary knows the threshold values for each individual in the population. While the implications of surveillance on targeting and unraveling might not be surprising, we highlight here the direct implications of the information that Web 2.0 technologies provide about people -- including election officials -- as well as the organization and delivery of harassment on and off these platforms. 

With information provided by online collection from public profiles and previous messages and posts, the adversary can formulate an optimum strategy. At each time step, this strategy is to target the maximum threshold of officials such that after distributing the net harassment $H$ among the targeted group, they all leave --- the highest threshold of officials who can be included among harassment targets and still leave the population even after the net harassment $H$ is distributed among the targets. Our model provides a framework for understanding how targeted harassment can have cascading indirect effects. If we denote the maximum threshold of people targeted at time $t$ by $\theta_t$, then at time zero, $\theta_0 = 0$, at time one, all officials with a threshold less than $\theta_1 = \max\{\theta \geq \theta_0 = 0 : \theta \leq H/(F(\theta) - F(0)) = H/F(\theta)\}$ will be targeted and removed. At any following time step, $t+1$, the adversary will target all officials whose thresholds are less than $\theta_{t+1}$, where $\theta_{t+1}$ is the largest number $\vartheta \geq \theta_t$ such that $\vartheta \leq H/(F(\vartheta) - F(\theta_{t}))$:
\begin{align}
    \theta_{t+1} = \max\{ \vartheta \geq \theta_{t}:  \vartheta \leq  H/(F(\vartheta) - F(\theta_{t}))\}.  \label{eq:model-3-sys-dynamics}
\end{align} In \cref{app:surviellence-and-tagetting} we show this dynamic is equivalent to \eqref{eq:model-3-sys-dynamics-eq}:
\begin{align}
    \theta_{t+1} = \frac{H}{F(\theta_{t+1}) - F(\theta_t)},  \label{eq:model-3-sys-dynamics-eq}
\end{align} and as long as $F$ is a continuous distribution everyone in the community will leave after a finite time, i.e., catastrophic unravelling of the election system becomes unavoidable in this case. One can also consider intermediate settings where an adversary only has noisy information about the thresholds and thus cannot target them precisely. In that case, the target set may include people whose thresholds are too large to be affected. We expect the conclusion that any continuous distribution will eventually be unravelled should remain true as long as the noise level is small enough. The largest noise level that allows for total unravelling can be calculated as a function of the distribution and could depend on distributional features such as variance.\footnote{This goes beyond the current work, but could be explored with constraints on the threshold distributions that could be calibrated by surveys of election officials.}  

% \begin{theorem} \label{thm:model-3-collapse}
% The resilience of Model \eqref{eq:model-3-sys-dynamics} is $\mathcal{R}(F) = 0$.
% \end{theorem}

% Both of the proofs are shown in Section \eqref{sec:model3}.

\section{Costly Actions and Optimized Attack Policies}\label{sec:plan}
Here we assume that the attacker pays a cost $c \cdot h_t$ to incur harassment $h_t$ at time $t$ and enjoys a reward $r_t = (p_{t+1} - p_{t}) - c \cdot h_t$ equal to the proportion of people leaving the system at time $t$ less the cost of harassment applied at time $t$. The dynamics of attrition in this case is given by $\displaystyle p_{t+1} = F({h_t}/({1 - p_t}))$. To investigate the optimal policy of the attacker, we can model the attacker problem as an infinite-horizon, discounted-payoff, multistage decision process, with input $h_t$ and state transition dynamics: $\displaystyle p_{t+1} = T(p_t,h_t) =  F({h_t}/({1 - p_t}))$. The utility of the attacker in this case is given by:
\begin{align*}
  u_\pi & = \sum_{t=0}^\infty \delta^t \, r_t \\ 
        & = \sum_{t=0}^\infty \delta^t (p_{t+1} - p_{t} - c \cdot h_t)\\ 
        & = \sum_{t=0}^\infty \delta^t (F(\frac{h_t}{1 - p_t}) - p_{t} - c \cdot h_t),
\end{align*} under the policy $\pi = [h_0, h_1, h_2, \cdots]$, with initial condition $p_0 = 0$, and discount factor $0\leq\delta\leq1$. Here the attacker is interested in causing as much attrition as soon as possible, while limiting their incurred cost. If the maximum applicable harassment at each time step $t$ is bounded by $H$, then the rewards are also bounded ($|r_t | \leq 2 + c H =: M$ ), and so is the utility $|u_\pi| = \sum_{t=0}^\infty \delta^t \, r_t \leq {M}/{(1 - \delta)}$ for any policy $\pi$. Defining  
\begin{align}
    R(p,h)  = F(\frac{h}{1 - p}) - p - c \cdot h, \mbox{ for } 0\leq p \leq 1 \mbox{ and } 0\leq h \leq H, \label{eq:reward}
\end{align} we can rewrite the attacker's utility as $u_\pi = \sum_{0}^{\infty}\delta^t R(p_t,h_t)$, which is uniquely maximized by a time-invariant, Markovian policy $\pi^*$ that gives the optimal harassment at any time $t$ as a function of the current state $p_t$: $h_t = \pi^*(p_t)$. The optimal policy $\pi^*$ is given by the following Bellman equation \citep{bertsekas2012dynamic}: 
\begin{align}
    \pi^*(p) = \argmax_{0 \leq h \leq H}\{R(p,h)+\delta V(F(h/(1-p)))\}, \label{eq:optimal-policy}
\end{align} where $V(\cdot)$ is the value function given by the following functional equation:
\begin{align*}
    V(p) = \max_{0 \leq h \leq H}\{R(p,h)+\delta V(F(h/(1-p)))\}, \mbox{ for all } 0 \leq p \leq 1,
\end{align*} which can be approximated iteratively. \newline

  % Therefore, we can use the Value Iteration Algorithm to compute optimal attack plans by solving the Hamilton-Jacobi-Bellman equation. \newline

% \begin{algorithm}[h]
% 	\caption{Value Iteration}
% 	\label{alg:Value Iteration}
% 	\KwIn{Distribution Function $F$ of the Thresholds $\theta \sim F$}
% 	\KwOut{Optimal Policy $\pi = [\mu_0^*, \mu_1^*, \cdots, \mu_t^*, \cdots]$}  
% 	\BlankLine

%         {\color{red} For typical setting: $\mu = \mu_0 = \mu_1 = \mu_2 = \cdots = \mu_t = \cdots$ and $h_t = \mu_t(p_t)$}
        
% 	Initialize $u^{(0)} = 0$ or randomly;
	
% 	\While{\textnormal{not converged}}{
	
% 		Update $\displaystyle u^{(k+1)} = \max_{h} \Big( r_t(h) + \alpha \cdot u^{(k)} \Big)$; 

% 	}
	
% \end{algorithm}

% The dynamic $\displaystyle f(p_t, l_t = p_t - p_{t-1}, h_t) \Rightarrow p_t = F(\frac{h_t}{l_t})$

% It is a DP because the input and the change occur at time $t$, and it doesn't matter what state variable is?

% \subsubsection{Attack strategies subject to budget constraints}

% Given $F$ or $\widehat{F}$, solve the MDP 
% % with noise 

% PYTHON CODES TO DO

\noindent{\bf The value of information to the attacker.} The above optimal attack plan is computed as a function of the distribution $F$. When distribution $F$ is unknown but can be learned up to a precision at a cost, we can perform a value information analysis for the cost of learning $F$ more precisely.  Let $\widehat{F}$ be a noisy estimate of $F$, whose precision is measured by a distributional distance function, denoted by $\|F - \widehat{F}\|$. Let $u_{\pi^{\star}}$ be the optimal utility of the attacker given $F$, and let $u_{\hat{\pi}}$ be the utility when policy $\hat{\pi}$ is optimized for the estimate $\widehat{F}$. The value of information analysis in this case reduce to the choice of a function $f$ such that $\|F - \widehat{F}\| \leq \epsilon \Rightarrow \|u_{\pi^{\star}} - u_{\hat{\pi}}\| \leq f(\epsilon)$: the attacker would be willing to pay $f(\epsilon)$ to obtain an estimate distribution $\widehat{F}$ that is $\epsilon$-more precise. The derivative $f'(\epsilon)$ measures the price tag that the attacker is willing to pay per unit improvement of the precision locally at the $\epsilon$-level. \newline

\noindent{\bf Increasing the cost of harassment to the attackers.} The dynamic programming formulation of the attacker's optimal policy allows us to investigate the consequences of increasing the harassment cost to the attacker. Threats of legal action, deplatforming and selective bans against harassers, as well as providing special protections to election officials are all examples of ways in which harassment can be made more costly to the attacker. A natural questions is how large $c$ needs to be to completely disincentivize any harassment, i.e., $\pi^*(p) = 0$ for all $p$ in \cref{eq:optimal-policy}. Setting $\partial{R(0,h)}/\partial{h} \leq0$ in \cref{eq:reward} gives the following necessary condition: $c\geq\max_{0\leq h \leq H}f(h)$, where $f(\cdot)$ is the probability density function for the distribution of the thresholds. Going back to the examples in \cref{fig:distributions}, for both $F_1 \sim \mbox{Uniform}[0, 1]$ and $F_2 \sim \mbox{Exponential}(1)$ we need at least $c \geq 1$.

%\subsection{Simulation}
%Writing of simulation part here. Figures included..................................................................................................................................................................................................................%
%In the previous subsection, we proved that when the distribution of officers' threshold satisfies certain conditions, there exists the maximum harassment that the system can sustain and the corresponding proportion of officers who leave the network because of the harassment. To further verify the theorem, simulations of the system were performed to reflect a more realistic situation and the results were visualized and analyzed based on the results of the simulations.

%In this section, we take the exponential distribution as an example of the distribution of officers' thresholds. We choose three, five, and eight as the parameters of the exponential distribution. Initially, we assume that the number of officials in the community is 1,000. 
%%% Write the parameters, setup here

%To verify the previous theorem and the effects of the parameters of the distribution of officers' threshold, we separate the simulation process into two parts. In the first part, we change the parameter in the distribution
% Describe the first/second part of the simulation (and the result)

%%%%%Insert figures here!

%Subsection?
%%%%%%%%%%

\section{Interventions to Increase Resilience}\label{sec:intervention}

 The analysis above identifies a mechanism whereby micro-level heterogeneity interacts with sociotechnical shocks to produce macro-unraveling of election infrastructure. Having isolated these mechanisms and interactions, we can now model the interventions (e.g., increasing salary incentives or harassment protections to better retain election officials) as a costly shift in the distribution of thresholds. Recall \cref{prop:stochstic-ordering} that a stochastic ordering of the distribution of thresholds implies a similar ordering of their resilience. Hence, interventions that change the distribution $F$, making the thresholds on average large, should increase resilience $\mathcal{R}(F)$. We can use an optimal transport formulation of the intervention design problem to find the least cost modification of the thresholds from an initial $F$ to a target distribution $F^{\star}$, where $F^{\star}$ solves: 
\begin{equation} \label{OT1}
    \begin{split}
    \min_{\phi} \quad & W_p(F, \phi)  \\
     \text{s.t.} \quad & R(\phi) \geq R
    \end{split}
\end{equation} to ensure that a harassment $R$ is tolerable with the least cost $W_p(F, F^{\star})$ or  
\begin{equation} \label{OT2}
    \begin{split}
    \max_{\phi} \quad & R(\phi) \\
     \text{s.t.} \quad & W_p(F, \phi) \leq B
    \end{split}
\end{equation} to ensure that $F^{\star}$ is the most resilient distribution that is achievable with a budget of $B$. The cost function $W_p(F, \phi)$ in these formulations is the $L_p$ Wasserstein (earth mover) distance give by:
\begin{equation} \label{OT-Kantorovich}
    \begin{split}
    W_p(F_1, F_2) = & \min _{{\gamma \in \Gamma (F_1 ,F_2 )}}\int _{{{\mathbf  {R}}^{2}}}  \|x - y\|_p \,{\mathrm  {d}}\gamma (x,y)   \\
     & \text{s.t.} \quad  \int \gamma(x, y) dy = f_1(x), \\
     & \quad \quad \int \gamma(x, y) dx = f_2(y),
    \end{split}
\end{equation} which measures the least-cost effort of transporting thresholds in distribution $F_1$ (with density function $f_1$) to thresholds in distribution $F_2$ (with density function $f_2$); $\Gamma (F_1 ,F_2 )$ is the space of all joint distributions with $F_1$ and $F_2$ as their marginals. While the exact details of how much specific budgets for policy interventions will boost resilience depend on the contexts (i.e., the thresholds), our work here highlight how policy makers, civic leaders, and those interested in promoting robust election infrastructure can actively respond to new sociotechnical challenges and evidence of targeted harassment.   

\section{Discussion and Conclusions}\label{sec:conc}

In this article, we study the democratic resilience when civic networks are targeted
with harassment, for example, in the case of election administrators. Even if the
total harassment is constant and equally distributed, as some election officials
leave the network, those who remain in the system will experience increasing
harassment which can result in a cascading failure that unravels the civic network
(e.g., removing public support for a policy). This unravelling may be only partial
if the input harassment is small enough. We characterize resilience as the
least harassment that causes total unravelling and relate it to network heterogeneity
and distributional features. We also consider more sophisticated adversaries who
can target their harassment and show the resultant fragility of the system (total
unraveling) in the face of targeted attacks. New information communication technologies can make individual thresholds partially or perfectly observable, leading to new risks and opportunities. Interventions that increase individual thresholds can be optimized to achieve maximum systemic resilience. On the other hand, adversaries who surveil social networks can also target their interventions to cause maximum unraveling with a fixed harassment budget. In addition, information to inform disruptive targeting can interact with other factors, such as the ages of election officials and those with accumulated expertise \citep{BipartisanCenterReportData2024}, to accelerate dysfunction. Additional research beyond our current modeling will be needed to unpack these dynamics.

Beyond the elections context, harassment reduces market efficiency and disrupts our organizational ability to efficiently aggregate information and make decisions collectively; e.g., public health experts facing harassment during the COVID-19 pandemic were unable to effectively guide public opinion toward the best public health practices. Although protesting is a democratic right, harassment crosses lines -- especially if motivated by false rumors -- that can uselessly disrupt the wisdom of crowds and waste scarce social capital. Building democratic resilience against harassment, while protecting the right to protest, is essential to the survival of critical civic infrastructure in the new (mis)information age. \newline

\section*{Compliance with Ethical Standards}

MAR has served on the advisory committee of a vaccine confidence fund created by Meta and Merck, and some of his research has also been funded by Meta. MPC declares no competing interests. The research does not involve human participants and/or animals.  Although our work is purely theoretical in nature and does not -- to the best of our knowledge -- pose specific ethical challenges, the potential societal and policy impact of our work is acknowledged and thoroughly debated as the subject matter of the paper. \newline

% One of the authors has served on the advisory committee of a vaccine confidence fund created by Meta and Merck, and some of their research has also been funded by Meta. The other author declares no competing interests. The research does not involve human participants and/or animals.  Although our work is purely theoretical in nature and does not -- to the best of our knowledge -- pose specific ethical challenges, the potential societal and policy impact of our work is acknowledged and thoroughly debated as the subject matter of the paper. \newline

\backmatter

% \bmhead{Supplementary information}

% If your article has accompanying supplementary file/s please state so here. 

% Authors reporting data from electrophoretic gels and blots should supply the full unprocessed scans for key as part of their Supplementary information. This may be requested by the editorial team/s if it is missing.

% Please refer to Journal-level guidance for any specific requirements.

\noindent{\bf Acknowledgements.} 
 %Removed for anonymity. 
  M.A.R. was partially supported by NSF SaTC-2318844. The authors would like to thank the two anonymous reviewers, the participants at the 2023 IDeaS Conference, Natalie Scala, Dominic Bordelon, and Jacob Otto for their valuable discussions and feedback.

\begin{appendices}
\crefalias{section}{appendix}
% \section{Section title of first appendix}\label{secA1}

% An appendix contains supplementary information that is not an essential part of the text itself but which may be helpful in providing a more comprehensive understanding of the research problem or it is information that is too cumbersome to be included in the body of the paper.

\section{Monotonicity and Convergence}\label{app:monotonicity-convergence}
\begin{proof}
    We can show this by induction. The base case is true because the sequence starts with $p_0 = 0$ and $\displaystyle p_1 = F({H}/{(1-p_0)}) = F(H) \geq 0 = p_0$. To show the induction step, assume $p_{t} \geq p_{t-1}$, and note that for any cumulative distribution function $F$, we have ${\displaystyle p_{t+1} = F({H}/{(1-p_t)}) \geq F({H}/{(1-p_{t-1})}) = p_t }$. Therefore, $\{p_t\}$ increases monotonically. Since $p_t$ is bounded by $1$, by the monotone convergence theorem, the sequence $p_t$ converges to a limit.
\end{proof}

\section{Resilience and Quantile Function}\label{app:quantile} 
\begin{proof}
        Define $\displaystyle g_H(x) = x - F({H}/({1-x}))$, and note that $x = P_{\infty}(H) =1$ is always a root: $g(1; H) = 0$. Moreover, $-1 \leq g(0; H) = -F(H) \leq 0$. If $F(H) = 0$, then $p_{\infty}(H) = 0$. Hence, to find $\mathcal{R}(F) = \inf \{H>0 \mid p_{\infty}(H) = 1\}$, we focus on values of $H$ such that $F(H) > 0$ and $g_{H}(0) = -F(H) <0$. To have $p_{\infty}(H) = 1$, it is necessary and sufficient that $g_{H}(x)$ has no roots in the interval $[0, 1)$. Setting $g_{H}(x) =  x - F({H}/({1-x})) < 0, \forall x \in (0, 1)$, applying $F^{-1} = Q$ to $x < F({H}/({1-x}))$ and using \cref{def:res} we get $\mathcal{R}(F) = \inf\{H \mid (1-p)Q(p) < H\} = \sup_{p \in (0,1)} (1-p)Q(p)$.
\end{proof}

\section{A Stochastic Ordering} \label{app:stochastic-ordering} 
\begin{proof}
     Note that $F_{1}$ and $F_{2}$ are both increasing functions, and we know that $F_{1}(x)\leq F_{2}(x)$ for all $x$. Because $F_{1}^{-1}$  is increasing, $F_{1}^{-1}(F_{1}(x)) \leq  F_{1}^{-1}(F_{2}(x))$ and since $F_{1}^{-1}(F_{1}(x))=x=F_{2}^{-1}(F_2(x))$, we get $ F_{2}^{-1}(F_{2}(x)) \leq F_{1}^{-1}(F_{2}(x))$ or $ F_{2}^{-1}(y) \leq F_{1}^{-1}(y)$, where $y=F_{2}(x)$, the inequality holding for all $y$. We know that $F_{1}$ stochastically dominating $F_{2}$ means that $F_{1}(x) \leq F_{2}(x)$ for all $x$, with a strict inequality at some $x$, and by monotonicity we showed that $F^{-1}_{2}(y)\leq F^{-1}_{1}(y)$ for all $y$, hence we can compare their supremum. That is, from the preceding equation we know that $(1-p)Q_{1}(p)=(1-p)F^{-1}_{1}(p)\geq (1-p)Q_{2}(p)=(1-p)F^{-1}_{2}(p)$ for all $p$, hence, $\mathcal {R}(F_1) = \sup _{p \in (0,1)} (1-p)Q_1(p) \ge \mathcal {R}(F_2) = \sup _{p \in (0,1)} (1-p)Q_2(p)$.
\end{proof}

% \section{Resilience and Infiltration}

\section{Surveillance and Targeting}\label{app:surviellence-and-tagetting} 

\begin{proof} We assume $F$ is a continuous distribution. Because $\theta_{t+1}$ is a feasible solution of \eqref{eq:model-3-sys-dynamics}, $\theta_{t+1} \leq {H}/{(F(\theta_{t+1}) - F(\theta_t))}$. Furthermore, if $\theta_{t+1} < {H}/{(F(\theta_{t+1}) - F(\theta_t))}$, then there must exist a $\delta > 0$ such that $\forall \, 0 \leq \epsilon < \delta: \theta_{t+1} + \epsilon < {H}/{(F(\theta_{t+1} + \epsilon) - F(\theta_t))}$ because of the continuity of $F$, which contradicts the optimality of $\theta_{t+1}$. Note also that $\theta_{t+1}$ forms an increasing sequence that is lower bounded as: $\theta_{t+1} \geq {H}/{(1 - F(\theta_t))}$, hence $\theta_t \to\infty$ or $F(\theta_t) \to 1$ as $t \to \infty$: the election system in this case will asymptotically unravel for all continuous distributions $F$.  
    
\end{proof}



\end{appendices}


\bibliography{main}% common bib file

\end{document}